\documentclass{article}
\usepackage{spconf,amsmath,graphicx,hyperref}
\usepackage{amssymb,booktabs,dsfont,multirow}
\usepackage[ruled,vlined]{algorithm2e}
\usepackage{bm}

\newcommand{\Vk}{\mathcal{V}_k}
\newcommand{\Ek}{\mathcal{E}_k}
\newcommand{\ind}{\mathds{1}}
\newcommand{\R}{\mathbb{R}}
\newcommand{\one}{\bm{1}}          
\newcommand{\inner}[2]{\langle #1, #2 \rangle}
\newcommand{\Ah}{\widehat{\mathcal{A}}} 
\usepackage{pifont}
\newcommand{\cmark}{\ding{51}}
\newcommand{\xmark}{\ding{55}}
\title{FDR-Controlled Variable Selection for Generalized Linear Models and Cox Regression with Virtual Dummies}
\name{Helena Mehler, Taulant Koka, Michael Muma\thanks{H. Mehler was funded by the German Research Foundation (DFG), project number 550090872. T. Koka and M. Muma have been funded by the ERC Starting Grant ScReeningData under grant number 101042407.
The authors gratefully acknowledge the computing time provided to them at the NHR Center NHR4CES at TU Darmstadt (project number p0020087). This is funded by the Federal Ministry of Research, Technology and Space, and the state governments participating on the basis of the resolutions of the GWK for national high performance computing at universities.
}}
\address{Robust Data Science Group, TU Darmstadt, Germany}
\begin{document}
\ninept
\maketitle
\begin{abstract}
In genomics, imaging and clinical studies, only a few of many candidate predictors are often nonlinearly associated with a response that may be, e.g. binary, categorical, a count or a censored event time. The Terminating-Random Experiments (T-Rex) selector is a scalable variable selection method that controls the false discovery rate (FDR) by letting synthetic null variables (dummies) compete with the real predictors. While the FDR control theory embraces more general settings, to date, the T-Rex selector has been specified only for linear models. We propose a memory-efficient selection procedure with FDR control for generalized linear models and Cox regression by extending the recently developed virtual dummy construction to score-based forward selection for Bernoulli, Poisson, multinomial and Cox responses. The virtual-dummy-based selection path remains equal in distribution to explicit
augmentation, so FDR control carries over under the same assumptions.
Simulations confirm this equivalence and the power gained by correct model
specification. Real-world applicability is illustrated on simulated genotypes and on cancer survival data.
\end{abstract}


\section{Introduction}
\label{sec:intro}
Genome-wide association studies \cite{abdellaoui202315}, single-cell sequencing \cite{townes2019feature}, cancer survival analysis \cite{bovelstad2007predicting} and {fMRI} brain decoding \cite{norman2006beyond} all require addressing the following task: among a large number of candidate predictors, identify the few that are truly associated with a response, which is often binary, categorical, a count or a censored event time. The number of candidates $p$ typically far exceeds the number of observations 
$n$, while only a small fraction is active, which leads to false positive selections that limit reproducibility and potentially lead to a costly follow-up experiment on an inactive variable. The main current paradigm is to control false discovery rate (FDR) \cite{benjamini1995controlling}, the expected proportion of false discoveries among the selected variables, while recovering as many active predictors as possible.

Classically, Benjamini-Hochberg and Benjamini-Yekutieli \cite{benjamini1995controlling,benjamini2001control} control the FDR using $p$-values, which are, however, not valid in a multivariate model with more predictors than observations. Model-X knockoffs \cite{barber2015controlling, candes2018panning} control the FDR by creating knockoff copies for each variable from an estimated model of the predictor distribution. However, the framework becomes prohibitively computationally demanding beyond a few tens of thousands of variables \cite{machkour2025terminating,koka2026virtual}.

The Terminating-Random Experiments (T-Rex) selector \cite{machkour2025terminating} scales linearly in the number of variables and allows for computing problems with millions of variables. It requires neither $p$-values nor a model of the predictor distribution: Instead, it estimates the FDR by evaluating the competition of synthetic null variables (dummies) with the real predictors in multiple early terminated forward selection runs. The recently developed virtual dummy (VD) construction of \cite{koka2026virtual} removes the memory bottleneck of storing large dummy matrices without changing the selection path or the FDR guarantee. Although the theory is more general, the construction has not yet been specified beyond the linear model, which can lead to a loss in statistical power for nonlinear models.

The contribution of this paper is a computationally efficient variable selection procedure with FDR control for generalized linear models and Cox regression. To this end, we extend the virtual dummy framework \cite{koka2026virtual} to score-based forward selection and instantiate it on orthogonal matching pursuit (OMP) with Bernoulli, Poisson, multinomial and Cox responses. The virtual dummies based extension preserves the equivalence to explicitly generated dummies.
In simulations, we confirm this equivalence and show that correct specification yields higher power than selection under a misspecified linear model. Promising results are obtained in two genomic applications.

\section{Preliminaries}
\label{sec:prelim}
We consider a variable selection problem with response vector $\bm y\in\R^n$ and predictor matrix $\bm X=(\bm x_1,\dots,\bm x_p)\in\R^{n\times p}$, whose columns are centered and $\ell_2$-normalized and hence lie in the centered subspace $H=\{\bm x\in\R^n:\one^\top\bm x=0\}$ of dimension $m=n-1$. We assume that the response $\bm y$ depends on the predictors only through the linear predictor $\bm\eta=b_0\one+\bm X\bm\beta$. In a generalized linear model, a link function $g$ relates the conditional mean of $\bm y$ to $\bm \eta$, whereas in the Cox model, $\bm \eta$ determines the hazard ratio. The coefficient vector $\bm\beta\in\R^p$ is sparse, and the goal is to recover its support, the active set $\mathcal{A}=\{j:\beta_j\neq0\}$. The selected set $\Ah$ should contain as many active predictors as possible, i.e., maximize the true positive rate (TPR), while keeping the FDR at or below a user-defined target level $\alpha$.

\subsection{The T-Rex Selector}\label{sec:trex}
The T-Rex selector \cite{machkour2025terminating} is a high-dimensional variable selection framework that controls a user-defined target FDR while maximizing the number of selected variables, by fusing the solutions of $B$ early terminated random experiments. In each experiment, $L$ dummies $\bm d_1,\ldots,\bm d_L$ are drawn i.i.d.\ from a standard normal distribution, independently of $(\bm X,\bm y)$, and collected in $\bm D$. A forward selection algorithm is run on $(\bm X\;\bm D)$ and $\bm y$, and the forward path is terminated once $T$ dummies have been selected. Let $\mathcal{C}_{b,L}(T)$ be the set of real variables selected when the $T$th dummy enters. The final set is determined by thresholding the relative occurrence
\begin{equation}
  \Phi_{T,L}(j)=\frac{1}{B}\sum_{b=1}^{B}\ind\{j\in\mathcal{C}_{b,L}(T)\}
  \label{eq:relocc}
\end{equation}
of each variable across all random experiments at a voting level $v\in[0.5,1)$, i.e., $\Ah(v,T)=\{j:\Phi_{T,L}(j)>v\}$, where $\ind\{\cdot\}$ denotes the indicator function and $(v,T)$ are calibrated to maximize $|\Ah_L(v,T)|$ such that an estimate $\widehat{\mathrm{FDP}}(v,T,L)$ of the FDP stays below the target level $\alpha$ \cite{machkour2025terminating}.

While the computational complexity is linear in $n$ and $p$ \cite{machkour2025terminating}, its bottleneck in very large settings is the memory required for the dummies: each random experiment stores an $n\times L$ matrix with $L\geq p$.

\subsection{Virtual Dummies}
\label{sec:vd}
The virtual dummy framework of \cite{koka2026virtual} removes the memory bottleneck by leveraging the fact that many forward selection procedures do not require the $n$ coordinates of a dummy, but only its projections onto an adaptively growing subspace. In this way, it completely avoids augmenting and storing a dummy matrix.
 
More specifically, let $\Vk$ be the subspace revealed after $k$ steps and $\Ek=\{\bm e_1,\ldots,\bm e_k\}$ an orthonormal basis of it, obtained by orthonormalizing the revealed directions in the order of selection. Each dummy $\bm d_\ell$ is then stored as the growing collection of scalars $\alpha_{i\ell}:=\inner{\bm d_\ell}{\bm e_i}$, $i=1,\ldots,k$, instead of an $n$-vector. Since the dummies are uniformly distributed on the unit sphere $\mathbb{S}_H$ in $H$ after centering and normalization, these coefficients can be drawn one at a time from their exact conditional law given the selection and projection histories, by adaptive stick-breaking: with $R_{0\ell}^2=1$, a Rademacher sign $S_{k\ell}$ and $U_{k\ell}\sim\mathrm{Beta}(\tfrac12,\tfrac{m-k}{2})$,
\begin{equation}
  \alpha_{k\ell}=S_{k\ell}R_{k-1,\ell}\sqrt{U_{k\ell}},
  \qquad
  R_{k,\ell}^2=R_{k-1,\ell}^2\bigl(1-U_{k\ell}\bigr),
  \label{eq:stick}
\end{equation}
where $R_{k,\ell}$ is the part of dummy $\ell$ that is not yet revealed. A full $n$-dimensional vector is realized only for a dummy that is actually selected. Drawn this way, the virtual and the augmented selection path are identical in distribution, so that the FDR control of \cite{machkour2025terminating} carries over to the virtual dummy framework \cite{koka2026virtual}. 
\section{Extension to GLMs and Cox Regression}
\label{sec:vdglm}
We now propose an FDR-controlled selector for GLMs and Cox regression. Section~\ref{sec:afs} presents our forward selection algorithm, ranking candidates by the score of a maximum-likelihood (ML) fit. Section~\ref{sec:mod} adapts the virtual dummy construction to this selector.

\subsection{Score-Based Forward Selection}
\label{sec:afs}
Both the virtual dummies and the data augmented T-Rex selector algorithms are compatible with a large class of forward selection algorithms (see \cite{koka2026virtual}), for example OMP \cite{pati1993orthogonal}. At step $k$, OMP selects the candidate most correlated with the current residual $\bm r_{k-1}$, i.e., $j_k^\star=\arg\max_j|\inner{\bm x_j}{\bm r_{k-1}}|$, adds it to the selected set $\Ah_k$ and refits least squares on $\bm X_{\Ah_k}$.
For the canonical-link GLMs considered here, the residual is replaced by the score contribution
\begin{equation}
  \bm s_k=\bm y-\bm\mu_k,\qquad \bm\mu_k=g^{-1}(\bm\eta_k),\quad
  \bm\eta_k=b_0^{(k)}\one+\bm X_{\Ah_k}\bm\beta^{(k)},
  \label{eq:score}
\end{equation}
where $(b_0^{(k)},\bm\beta^{(k)})$ solves the ML problem of the GLM on $\Ah_k$, and the selection rule reads $j_k^\star=\arg\max_j|\inner{\bm x_j}{\bm s_{k-1}}|$.

Forward selection procedures that take smaller steps can be used in place of OMP, such as adaptive forward stepwise (AFS) \cite{zhang2026adaptive}: it advances the coefficients only a fraction  $\rho\in(0,1]$ towards the refit and otherwise ranks the candidates in the same way.
 
In addition to standard GLMs such as Poisson or Bernoulli, we implement a multinomial response and, beyond the GLM class, the semi-parametric Cox model; Table~\ref{tab:models} lists the score $\bm s_k$ for each model. 
The multinomial fit returns one score per class: with $C$ classes and one held out as reference, the softmax fit gives $C-1$ scores $\bm s_k^{(c)}$. The $C$ class residuals satisfy $\sum_{c=0}^{C-1}\bm s_k^{(c)}=\bm 0$, so the held-out score is determined by the others and a candidate is ranked by the $\ell_2$ norm of its inner products with all $C$ of them, which is invariant to the reference class. Cox regression maximizes a partial likelihood, and its score is the martingale residual $\bm s_k=\bm\delta-e^{\bm\eta_k}\hat\Lambda_0(t)$, where $\bm\delta$ is the event indicator and $\hat\Lambda_0$ the Breslow estimate of the baseline cumulative hazard.

\begin{table}[t]
\centering
\caption{Score $\bm s$ per response model, with
$\bm\eta=b_0\one+\bm X_{\Ah}\bm\beta$ and $\mathrm{Cat}$ the categorical
distribution over the $C$ classes.}
\label{tab:models}
\footnotesize
\setlength{\tabcolsep}{6pt}
\begin{tabular}{@{}lll@{}}
\toprule
Model & $y_i\sim$ & Score $\bm s$ \\
\midrule
Gaussian    & $\mathcal{N}(\eta_i,\sigma^2)$              & $\bm y-\bm X_{\Ah}\bm\beta$ \\
Bernoulli   & $\mathrm{Bern}(\sigma(\eta_i))$             & $\bm y-\sigma(\bm\eta)$ \\
Poisson     & $\mathrm{Pois}(e^{\eta_i})$                 & $\bm y-e^{\bm\eta}$ \\
Multinomial & $\mathrm{Cat}(\mathrm{softmax}\,\bm\eta_i)$ & $\bm y^{(c)}-\mathrm{softmax}^{(c)}\bm\eta$ \\
Cox & $\lambda_0(t)\,e^{\eta_i}$ & $\bm\delta-e^{\bm\eta}\hat\Lambda_0(t)$ \\
\bottomrule
\end{tabular}
\end{table}

\subsection{Virtual Dummies for Score-Based Selection}
\label{sec:mod}
We now adapt the virtual dummy construction of Section~\ref{sec:vd} for the
score ranking of Section~\ref{sec:afs}. At each step, the candidates are ranked by
their inner product with the current score $\bm s_{k-1}$, which is
$\inner{\bm x_j}{\bm s_{k-1}}$ for a real predictor. A dummy enters this ranking
only through its projections onto the previous scores
$\bm s_0,\ldots,\bm s_{k-1}$. Orthonormalizing these in the order of selection
gives the basis $\Ek$, whose span $\Vk$ contains the
current score at every step.

Each dummy can be split into a revealed and an unrevealed part,
\begin{equation}
  \bm d_\ell=\sum_{i\leq k}\alpha_{i\ell}\bm e_i+\bm d^{\perp}_{\ell,k},
  \qquad \bm d^{\perp}_{\ell,k}\in H\cap\Vk^{\perp},
  \label{eq:split}
\end{equation}
and since $\bm s_{k-1}\in\Vk$, the unrevealed part drops out of the inner
product, which reduces to
\begin{equation}
\inner{\bm d_\ell}{\bm s_{k-1}}=\sum_{i\leq k}\alpha_{i\ell}\inner{\bm e_i}{\bm s_{k-1}}.
\label{eq:dummyscore}
\end{equation}
The inner product is therefore computable from the $\alpha_{i\ell}$ alone.

Then, the candidate with the largest absolute score is selected. If it is a dummy, it has to be realized first: its unrevealed part $\bm d^{\perp}_{\ell,k}$ is drawn from its conditional law on $H\cap\Vk^{\perp}$ and added to the revealed part $\sum_{i\leq k}\alpha_{i\ell}\bm e_i$. The GLM is then refit by ML on the enlarged active set, which yields the new score. The score is centered and orthonormalized into the basis,
\begin{equation}
  \bm u_k=\bm s_k-\sum_{i=1}^{k}\inner{\bm s_k}{\bm e_i}\bm e_i,
  \qquad
  \bm e_{k+1}=\frac{\bm u_k}{\|\bm u_k\|_2},
  \label{eq:basisupdate}
\end{equation}
where the centering keeps $\bm e_{k+1}$ in $H$, on whose unit sphere the dummy coefficients live. One new coefficient $\alpha_{k+1,\ell}$ is then drawn from \eqref{eq:stick} for every dummy that is still unrealized.

For a multinomial response the refit returns $C-1$ scores, which all enter the ranking and therefore all have to lie in $\Vk$. Update \eqref{eq:basisupdate} is applied to each in turn, revealing up to $C-1$ directions per step, with one coefficient from \eqref{eq:stick} per direction and unrealized dummy, and the sums in \eqref{eq:split} and \eqref{eq:dummyscore} run over all of them.

The equivalence to explicit augmentation of the virtual dummies algorithm rests on two conditions: the dummies are drawn independently of $(\bm X,\bm y)$ with coefficients from \eqref{eq:stick}, and the selection rule and each new basis vector depend on the dummies only through the revealed coefficients and the dummies already realized. In this work, the first condition is unchanged. The second holds since the ML refit uses only $\bm y$ and the active columns to produce $\bm s_k$ and $\bm e_{k+1}$, and since $\bm s_{k-1}\in\Vk$, so every dummy correlation follows from \eqref{eq:dummyscore}; for a multinomial response, this applies to each class score and therefore to the norm over classes. 
Hence, by Theorem~1 and Corollary~2 of \cite{koka2026virtual}, the
proposed selector is equal in distribution to explicit augmentation and
inherits the FDR guarantee of \cite[Theorem~1]{machkour2025terminating}
under the original assumptions, notably mutual independence of the null
predictors and their independence from the active ones.
Algorithm~\ref{alg:vdglm} summarizes the proposed algorithm.

\begin{algorithm}[t]
\DontPrintSemicolon
\SetKwInOut{Require}{Require}
\caption{VD-T-Rex for GLMs and Cox}
\label{alg:vdglm}
\footnotesize
\Require{$\bm X=(\bm x_1,\dots,\bm x_p)\subset H$, $\bm y$, link $g$, $L$, $B$, $T_{\mathrm{max}}$, $\alpha$}
$T\leftarrow0$;\quad $\Ah_0,\mathcal{C}_{b,L}\leftarrow \varnothing$;\quad $k\leftarrow1$\quad $\forall b$\;
$\bm e_1\leftarrow\bm s_0/\|\bm s_0\|_2$;\quad
draw $\alpha_{1\ell}$, $\ell=1,\dots,L$ \tcp*{\eqref{eq:stick}}
\While{$T<T_{\mathrm{max}}$ \textbf{and} $\exists v\in[0.5,1):\widehat{\mathrm{FDP}}(v,T,L)\leq\alpha$}{
  $T\leftarrow T+1$\;
  \For{$b=1$ \KwTo $B$}{
    \While{$|\Ah_{k-1}\setminus\mathcal{C}_{b,L}|<T$}{
      $c_j\leftarrow|\inner{\bm x_j}{\bm s_{k-1}}|$\;
      $c_{p+\ell}\leftarrow|\sum_{i\leq k}\alpha_{i\ell}\inner{\bm e_i}{\bm s_{k-1}}|$\;
      $j^\star\leftarrow\arg\max_{j\notin\Ah_{k-1}}c_j$;\quad
      $\Ah_k\leftarrow\Ah_{k-1}\cup\{j^\star\}$\;
      \lIf{$j^\star$ is a dummy}{realize $\bm d_{\ell^\star}$ \tcp*[f]{\eqref{eq:split}}}
      \lElse{$\mathcal{C}_{b,L}\leftarrow\mathcal{C}_{b,L}\cup\{j^\star\}$}
      $\bm s_k\leftarrow\mathrm{ML\text{-}fit}(\bm y,\bm X_{\Ah_k})$ \tcp*{\eqref{eq:score}}
      $\bm e_{k+1}\leftarrow$ orthonormalize $\bm s_k$ against $\Ek$
      \tcp*{\eqref{eq:basisupdate}}
      draw $\alpha_{k+1,\ell}$ for every unrealized dummy\;
      $k\leftarrow k+1$\;
    }
  }
  $\Phi_{T,L}(j)\leftarrow\frac1B\sum_{b}\ind\{j\in\mathcal{C}_{b,L}\}$ \tcp*{\eqref{eq:relocc}}
  compute $\widehat{\mathrm{FDP}}(v,T,L)$ from $\Phi_{t,L}$, $t\leq T$, as in \cite{machkour2025terminating}\;
}
$(v^\star,T^\star)\leftarrow\arg\max_{v,T}|\Ah_L(v,T)|$ s.t.\ $\widehat{\mathrm{FDP}}(v,T,L)\leq\alpha$\;
\KwRet{$\Ah^\star=\{j:\Phi_{T^\star,L}(j)>v^\star\}$}
\end{algorithm}

\section{Experiments}
\label{sec:exp}

We evaluate the proposed selector in three steps. First, we verify the finite-sample equivalence of virtual and augmented dummies for GLMs and quantify the loss in power incurred under a misspecified Gaussian model. We then compare their memory use and runtime, and demonstrate real-world applicability in two genomic applications. Performance is assessed using the average false discovery proportion (FDP) and the true positive proportion (TPP), the empirical counterparts of FDR and TPR. Code: \href{https://github.com/helenamehler/virtual-dummies-glm.git}{https://github.com/helenamehler/virtual-dummies-glm.git}

\subsection{Equivalence and Misspecification}\label{sec:equiv} We consider a simulation setup with $n=300$ observations and $p=1000$ predictors, drawn i.i.d.\ standard normal, centered and $\ell_2$-normalized. The active set has size $a=10$ with random signs and a common magnitude, so that the linear predictor attains a target signal-to-noise ratio $\mathrm{SNR}=\mathrm{Var}(\mathbb{E}[y_i\mid\eta_i])/\mathbb{E}[\mathrm{Var}(y_i\mid\eta_i)]$, which is the classical $\mathrm{Var}(\bm X\bm\beta)/\sigma^2$ in the Gaussian case and, for the Cox model, is taken on the latent log-time scale as $\mathrm{Var}(\bm\eta)/(\pi^2/6)$. Responses follow the four models of Table~\ref{tab:models}: Bernoulli at prevalence $1/2$, Poisson at mean count $2$, multinomial with $C=4$ classes, each active predictor acting on one class, and Cox with $40\%$ censoring and exponential baseline hazard.

We run T-Rex with $B=20$ random experiments, $L=5p$ dummies and target FDR $\alpha=0.1$, and compare four methods. VD-GLM is the proposed selector, instantiated with the generating model, and AD-GLM its explicitly augmented counterpart. VD-Gaussian is the misspecified baseline using OMP with the linear model's least-squares residual. We also compare against the model-X knockoff filter \cite{candes2018panning}, with second-order Gaussian knockoffs and a coefficient-difference statistic from a Lasso fitted under the generating model. We average over $500$ replicates.

\begin{figure*}[t]
\centering
\setlength{\abovecaptionskip}{4pt}

\includegraphics[width=0.95\textwidth,trim=0 8 0 7,clip]{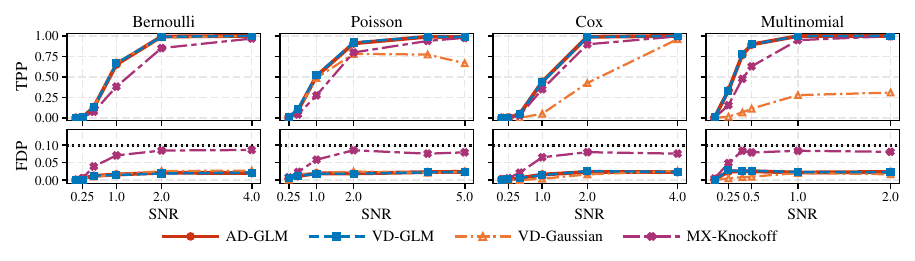}
\caption{TPP and FDP against the SNR, means over $500$ paired replicates,
$\alpha=0.1$ (dotted); setup in Sec.~\ref{sec:equiv}. Monte Carlo standard
errors of the AD-GLM, VD-GLM and VD-Gaussian curves are at most $0.003$ in FDP and
$0.013$ in TPP. AD-GLM and VD-GLM differ by at most $0.004$ and $0.017$.}
\label{fig:main}
\vspace{-14pt}
\end{figure*}

Figure~\ref{fig:main} provides empirical support that the GLM selector inherits the equivalence of \cite{koka2026virtual}: AD-GLM and VD-GLM coincide in FDP and TPP over the entire grid, and both control the FDR well below the target level. The cost of misspecification is highest for the multinomial response: VD-Gaussian stays at a TPP of $0.3$, while VD-Multinomial recovers nearly all active predictors. Under the Cox model the gap closes only at the highest SNR, and for Poisson counts it widens as the SNR grows. Only for the Bernoulli response is the loss in power negligible. The knockoff filter controls the FDP as well, but reaches a lower TPP than VD-GLM across all models over the whole grid.
\begin{table}[t]
\centering
\caption{HAPNEST disease classification, $n=10\,000$, $p=31\,037$,
$C=4$, $a=12$, $\alpha=0.1$, $B=20$, $L=5p$, $100$ replicates ($20$ for knockoffs). $^\dagger$Targets the FDR of marginal hypotheses}
\label{tab:hapnest_mn}
\footnotesize
\setlength{\tabcolsep}{4pt}
\begin{tabular}{@{}lcccc@{}}
\toprule
Method & $\overline{\mathrm{FDP}}\leq\alpha$ & $\overline{\mathrm{FDP}}$
& $\overline{\mathrm{TPP}}$ & $\overline{|\Ah|}$ \\
\midrule
\textbf{VD-Multinomial}  & \cmark & \textbf{0.048} & \textbf{0.453} & 5.8 \\
VD-Bernoulli, OvR        & \cmark & 0.0 & 0.0 & 0.0 \\
VD-Gaussian on labels    & \cmark & 0.018 & 0.181 & 2.2 \\
\midrule
Model-X Knockoffs       & \cmark & 0.0 & 0.0 & 0.0 \\
Multinomial Score + BH & \xmark$^\dagger$ & 0.279 & 0.579 & 10.3 \\
Multinomial Score + BY & \xmark$^\dagger$ & 0.121 & 0.474 & 6.7 \\
\bottomrule
\end{tabular}
\end{table}
\subsection{Memory and Runtime}
\label{sec:memory}
We compare the memory footprint and runtime of AD-GLM and VD-GLM at $n=10^4$, $p=10^5$ and $L=5p$, with each run executed in a separate single-threaded process, terminated once $T=10$ dummies have entered the active set, and report averages over 20 replicates. Storing $(\bm X,\bm y)$ alone takes $8.0$ GB, and AD-GLM adds $40.1$ GB of peak resident set size on top of it for the explicit dummy block. For the Poisson, Bernoulli and Cox instantiations, VD-GLM adds only around $205$ MB, and is $2.2$ to $2.5\times$ faster. The multinomial score has $C-1$ components, so the projection block grows accordingly to $411$ MB, still two orders of magnitude below AD-Multinomial, at a $2.1\times$ speedup. Absolute VD-GLM runtimes range from about $25$ s to $37$ s, against $58$ s to $89$ s for AD-GLM. The model-X knockoff method exceeded the wall-time limit of $36\,$h.

\subsection{Application 1: Disease Subtype Classification}
\label{sec:app2}

We first apply the multinomial instantiation of the proposed selector to a disease classification task on genotype data simulated with HAPNEST \cite{wharrie2023hapnest}, which reproduces the linkage disequilibrium (LD) structure and minor allele frequency spectrum of real human populations. We adopt the cohort and preprocessing of the small-scale setting of \cite{koka2026virtual}: quality control and LD pruning at $|\rho|>0.7$ leave $n=10\,000$ individuals and $p=31\,037$ SNPs on chromosome~1, with residual LD. The columns are centered and $\ell_2$-normalized.

The response has $C = 4$ classes, with the healthy controls as reference class. The other three are disease states, for instance cancer stages or stroke subtypes. Labels are drawn from a softmax over $a=12$ causal SNPs under a multiplicative relative risk model, with class prevalences $(\tfrac12,\tfrac16,\tfrac16,\tfrac16)$. Four loci are shared and raise the risk of every subtype, $\mathrm{RR}_j\sim\mathrm{Unif}(1.15,1.30)$, the other eight act on a single subtype, $\mathrm{RR}_j\sim\mathrm{Unif}(1.20,1.42)$. A SNP counts as active if it affects at least one class.

We run VD-Multinomial against two alternative specifications of the same selector, VD-Bernoulli one-vs-rest at level $\alpha/(C-1)$
and VD-Gaussian on the integer labels. The model-X knockoff filter \cite{candes2018panning} is instantiated with a multinomial model. Its runtime restricts this evaluation to $20$ replicates. As a marginal benchmark we use the multinomial score test, a generalization of the Cochran--Armitage trend test \cite{cochran1954some,armitage1955tests}, corrected by the Benjamini-Hochberg (BH) and the Benjamini-Yekutieli (BY) procedure \cite{benjamini1995controlling,benjamini2001control}.

Table~\ref{tab:hapnest_mn} shows that VD-Multinomial attains the highest power among the methods keeping the FDP below the target. VD-Gaussian reaches less than half that power and assumes a class ordering, while the one-vs-rest and the knockoff filter select nothing at all. The latter is also by far the most expensive methodology: $23.4\,$h per replicate, around $200$ times the $6.5\,$min of VD-Multinomial. The marginal procedures largely exceed the target FDP, as they test $\gamma_j=0$ for SNP $j$ fitted alone rather than $\beta_j=0$ in the joint model.

\subsection{Application 2: Survival in Clear-Cell Renal Cell Carcinoma}
\label{sec:kirc}
Finally, we apply the Cox instantiation of the proposed selector to gene expression and survival time data of clear cell renal cell carcinoma from The Cancer Genome Atlas (TCGA-KIRC) \cite{cancer2013comprehensive,goldman2020visualizing}. Starting from the gene-level RNA-seq counts of the TCGA-KIRC cohort, we keep primary tumor samples and protein-coding genes, discard genes with a total count below $10$ and patients with zero follow-up time, normalize by median-of-ratios size factors and apply a variance-stabilizing $\log_2$ transform. This leaves $p=19\,620$ candidate genes and $n=531$ patients, of which $175$ died during the study while the remaining survival times are right censored.

As benchmark methods, we consider marginal Cox score tests followed by BH and BY \cite{benjamini1995controlling,benjamini2001control}, the model-X knockoff method with a Cox Lasso statistic \cite{li2023coxknockoff}, and the cross-validated Cox Lasso \cite{simon2011regularization} at the one-standard-error criterion. For the proposed selector, we run VD-AFS-Cox, i.e. AFS with $\rho=0.1$ in place of OMP, see Sec.~\ref{sec:afs}.

\begin{table}[t]
\centering
\caption{TCGA-KIRC, $n=531$, $p=19\,620$, $175$ events, 
$\alpha=0.1$; VD runs use $B=20$, $\rho=0.1$, $L=5p$.}
\label{tab:kirc}
\footnotesize
\setlength{\tabcolsep}{4pt}
\begin{tabular}{@{}llr@{}}
\toprule
Method & Error criterion & $|\widehat{\mathcal{A}}|$ \\
\midrule
Marginal Cox + BH        & FDR, marginal  & 10\,837 \\
Marginal Cox + BY        & FDR, marginal  & 7001 \\
Cox Lasso, CV            & none           & 28 \\
\textbf{VD-AFS-Cox}      & FDR, joint     & \textbf{8} \\
VD-AFS-Gaussian on $\log t$ & FDR, joint  & 1 \\
Cox Knockoffs            & FDR, joint     & 0 \\
\bottomrule
\end{tabular}
\end{table}

Table~\ref{tab:kirc} lists the number of selected genes at $\alpha=0.1$. BH and BY target marginal associations and select several thousand genes. The Cox Lasso selects $28$, but its penalty is tuned by a cross-validated C-index and provides no FDR control. The knockoff filter selects no gene, and the linear model on $\log t$ selects one, against eight for VD-AFS-Cox. The number of selected genes therefore reflects the error criterion of each method, with the largest sets returned by procedures that either do not control FDR or control the FDR of marginal, rather than joint conditional, hypotheses.
The misspecified linear model on $\log t$ selects fewer genes.

Five of the eight selected genes have previously been linked to clear cell renal cell carcinoma: `DONSON' \cite{kluemper2020downstream}, `ZIC2' \cite{wu2021zic2}, `SLC16A12' \cite{mei2019decreased}, `SORBS2' \cite{lv2020rna} and `BARX1' \cite{sun2021transcription}; the remaining three are `HOXA2', `GPR78' and `SOWAHB'.

\section{Conclusion}
\label{sec:conclusion}
We extend the virtual dummy construction to score-based forward selection for Bernoulli, Poisson, multinomial and Cox responses, growing the basis along the centered score of the refitted model. Since the selector satisfies the conditions of the equivalence result in \cite{koka2026virtual}, its selection path is equal in distribution to explicit augmentation, and it inherits the FDR-control result of the T-Rex selector.

Simulations confirm this equivalence and show that misspecifying the response costs power, where the loss in TPR is most pronounced for multinomial and least for Bernoulli responses. On genomic data, the multinomial instantiation recovers more than twice as many causal SNPs as any competitor within the target FDR, and the Cox instantiation selects eight genes against one for the linear model on $\log t$.

Future work could include further selection procedures, among them robust forward selectors such as the LAD or Huber, or other path algorithms, such as greedy variants with variable deletion or more general pursuit rules as in \cite{scheidt2026greedy}.
\vfill\pagebreak
{\small
\bibliographystyle{IEEEbib}
\bibliography{strings,refs}}
\end{document}